\documentclass[aoas,nameyear,dvips]{arximspdf}
\usepackage{dcolumn}
\usepackage{graphicx}

\doi{10.1214/09-AOAS287}
\volume{4}
\issue{1}
\pubyear{2010}
\firstpage{484}
\lastpage{502}

\makeatletter
\newcolumntype{d}[1]{D{.}{.}{#1}}
\makeatother

\begin{document}
\begin{frontmatter}

\title{Downscaling extremes: A comparison of
extreme value distributions in point-source and gridded precipitation data}
\runtitle{Spatial scaling in climate data}

\begin{aug}
\author[A]{\fnms{Elizabeth C.} \snm{Mannshardt-Shamseldin}\corref{}\ead[label=e1]{elizabeth@stat.duke.edu}},
\author[B]{\fnms{Richard L.} \snm{Smith}\ead[label=e2]{rls@email.unc.edu}},
\author[C]{\fnms{Stephan R.} \snm{Sain}\ead[label=e3]{ssain@ucar.edu}},
\author[D]{\fnms{Linda O.} \snm{Mearns}\ead[label=e4]{lindam.ucar.edu}}
\and
\author[E]{\fnms{Daniel} \snm{Cooley}\ead[label=e5]{cooleyd@stat.colostate.edu}}
\runauthor{E. C. Mannshardt-Shamseldin et al.}
\affiliation{Duke University,
University of North Carolina at Chapel Hill,\\
The National Center for Atmospheric Research,\\
The National Center for Atmospheric Research,\\ and
Colorado State University}

\address[A]{E. C. Mannshardt-Shamseldin\\
Department of Statistical Science\\
Duke University\\
Durham, North Carolina 27708-025\\
USA\\
\printead{e1}}

\address[B]{R. L. Smith\\
Department of Statistics\\
\quad  and Operations Research\\
University of North Carolina\\
Chapel Hill, North Carolina 27599-3260\\
USA\\
\printead{e2}}

\address[C]{S. R. Sain\\
Institute for Mathematics Applied\\
\quad  to Geosciences\\
National Center\\
\quad  for Atmospheric Research\\
P.O. Box 3000\\
Boulder, Colorado 80307\\
USA\\
\printead{e3}}

\address[D]{L. O. Mearns\\
Institute for the Study of Society\\
\quad  and Environment\\
National Center for Atmospheric Research\\
P.O. Box 3000\\
Boulder, Colorado 80307\\
USA\\
\printead{e4}}

\address[E]{D. Cooley\\
Department of Statistics\\
Colorado State University\\
Ft.~Collins, Colorado 80523\\
USA\\
\printead{e5}}
\end{aug}

\received{\smonth{4} \syear{2008}}
\revised{\smonth{8} \syear{2009}}

%
\begin{abstract}
There is substantial empirical and climatological evidence that
precipitation extremes have become more extreme during the twentieth
century, and that this trend is likely to continue
as global warming becomes more intense. However, understanding these
issues is limited by a
fundamental issue of spatial scaling: most evidence of past trends
comes from rain gauge
data, whereas trends into the future are produced by climate models,
which rely on gridded aggregates.
To study this further, we fit the Generalized Extreme Value (GEV)
distribution to the
right tail of the distribution of both rain gauge and gridded events.
The results of this modeling exercise confirm
that return values computed from
rain gauge data are typically higher than those computed from gridded
data; however, the size of the difference is somewhat surprising, with
the rain gauge data exhibiting return values sometimes two or three
times that of the gridded data.
The main contribution of this paper is the development of a family of regression
relationships between the two sets of return values that also take
spatial variations into account. Based on these results, we now believe
it is possible to project
future changes in precipitation extremes at the point-location level
based on results from
climate models.
\end{abstract}

%

\end{frontmatter}

\section{Introduction}

There is great interest in understanding the behavior of the extremes
of weather and climate and the impacts of these extremes. Furthermore,
there is mounting evidence that, for example, precipitation extremes
have become even more extreme during the twentieth century, and that
this trend is likely to continue with continued global warming and
climate change
[see, e.g., Karl and Knight (\citeyear{KarlKnight1998}),
Zwiers and Kharin (\citeyear{ZwiersKharin1998}),
Groisman et al. (\citeyear{GroismanEtAl1999}),
Kharin and Zwiers (\citeyear{KharinZwiers2000}),
Meehl et al. (\citeyear{MeehlEtAl2000}),
Frich et al. (\citeyear{FrichEtAl2002}),
Kiktev et al. (\citeyear{KiktevEtAl2003}),
Hegerl et al. (\citeyear{HegerlEtAl2004}), and
Groisman et al. (\citeyear{GroismanEtAl2005})].
Future projections produced by global and regional climate models offer
a way to characterize any trends in extreme behavior. However, there is
the issue of \textit{spatial scaling} and how to compare the output of
climate models with historical data. Climate model data represent an
aggregate over a grid box, whereas historical data are collected from
rain gauges associated with monitoring stations at specific point locations.
In this work we seek to examine and quantify the relationship between
the extremes of gridded climatological data sets, such as reanalysis
data or climate model output, and observed point-level data from
weather stations. This relationship is used to develop a framework for
predicting point-level extreme behavior from future runs of climate
models. Essentially, we seek to construct a statistical model that
balances the following: (1)~exploiting the clear similarities in the
spatial patterns of extreme behavior from gridded and point-level data
sets and (2) accounting for the differences in the distributions of
extremes from the two types of data.

Our findings suggest that there is a family of regression
relationships between the two sets of return values that takes spatial
variation into account. Based on these results, we now believe it is
possible to project
future changes in precipitation extremes at the point-location level
based on results from
climate models.

The paper is organized as follows:
\noindent
First, we discuss the methodology for the statistical modeling of
extreme values and regression methods for analyzing the relationship
between gridded and point-level data.
Second, we look at gridded data results from a well-known reanalysis
(NCEP) and
from a climate model (CCSM). Spatial and temporal trends are
considered, and a comparison is made between NCEP and CCSM results.
To explore more fully the spatial trend using standard methods, we also
look at a comparison of the return values obtained through the modeled
regression relationship
with return values obtained through universal kriging over the unused stations.
Finally, a discussion of the results of this analysis and its
implications is presented.

\section{The data}

The point-level data were obtained from the National Climatic Data
Center (NCDC) and represent daily rainfall values at 5873
meteorological stations covering a period from 1950 to 1999.
\noindent
The reanalysis data are from the National Centers for Environmental
Prediction (NCEP) and cover a period from 1948 to 2003 on a 2.5$^{\circ}$
grid [Kalnay et al. (\citeyear{KalnayEtAl1996})], resulting in 288 grid cells. Precipitation
is determined by a numerical weather model in reanalysis data. It is
important to note that systematic model errors, due to incomplete
physical readings and grid resolution, can influence estimates obtained
from reanalysis data. The climate model output was obtained from two
runs of the National Center for Atmospheric Research's Community
Climate System Model (CCSM) which included a control run from 1970--1999
and a future projection from 2070--2099. The CCSM data were on a
1.4$^{\circ}$ grid (820 grid cells).
\noindent
The NCDC data were measured on a scale of tenths of a millimeter; the
NCEP and CCSM data were converted accordingly. Annual total rainfall
amounts were computed for each season: December, January, and February
(DJF); March, April, and May (MAM); June, July, and August (JJA); and
September, October, and November (SON).

\section{Methodology}

The statistical methodology adopted in this paper is essentially in two parts.
First, extreme value distributions are fitted to each data series (both
point-location
and gridded) to determine the 100-year return value for that series.
Second, regression
relationships are established between the point-location and gridded
return values,
primarily with the purpose of predicting the former from the latter.

In this section we briefly review extreme value theory, and then
explain how it
is applied to the present data sets. For further details the reader is
referred to overviews by Coles (\citeyear{Coles2001}) or Smith (\citeyear{Smith2003}).
Katz, Parlange and Naveau (\citeyear{KatzParlangeNaveau2002}) gave an excellent overview of the application of
extreme value methods in hydrology.

\subsection{The Generalized Extreme Value (GEV) distribution}\label{GEV}

Suppose $Y$ represents the annual maximum of daily precipitation in a given
series. The Generalized Extreme Value (GEV) distribution is defined by
the formula
%
\begin{equation}\label{E1}
\operatorname{Pr}\{Y\le y\} =
\exp \biggl\{- \biggl(1+\xi{y-\mu\over\psi} \biggr)_+^{-1/\xi} \biggr\},
\end{equation}
where $\mu$ is a location parameter, $\psi$ a scale parameter, and $\xi
$ is the extreme-value shape parameter; $\mu$ and $\xi$ can take any
value in $(-\infty,\infty)$ but $\psi$ has to be $>0$. The notation
$(\cdots)_+$ follows the convention $x_+=\max(x,0)$ and is intended to
signify that the range of the distribution is defined by $1+\xi{y-\mu
\over\psi}>0$. In other words, $y>\mu-{\psi\over\xi}$ when $\xi>0$,
$y<\mu-{\psi\over\xi}$ when $\xi<0$.

The distribution (\ref{E1}) encompasses the classical ``three types'' of
extreme value distributions [Fisher and Tippett (\citeyear{FisherTippett1928}), Gumbel (\citeyear{Gumbel1958})],
but in a form that facilitates parameter estimation through automated
techniques such as maximum likelihood. The ``three types'' correspond to
the cases $\xi>0$ (sometimes called the Fr\'echet type), $\xi<0$
(Weibull type), and $\xi=0$, which is interpreted as the limit case $\xi
\rightarrow0$ in (\ref{E1}),
%
\begin{equation}\label{E2}
\operatorname{Pr}\{Y\le y\} =
\exp \biggl\{-\exp \biggl(-{y-\mu\over\psi} \biggr) \biggr\},\qquad -\infty<y<\infty,
\end{equation}
widely known as the Gumbel distribution.

The \textit{n-year return value} is formally defined by setting (\ref{E1})
to $1-{1\over n}$; $y_n$ is then the solution to the resulting
equation. In practice, however, for large $n$, we have $1-{1\over
n}\approx e^{-1/n}$ and it is more convenient to define $y_n$ by the equation
\[
\biggl (1+\xi{y_n-\mu\over\psi} \biggr)^{-1/\xi} = {1\over n},
\]
which leads to the formula
\begin{equation}
y_n =
\cases{
\displaystyle\mu+\psi{n^\xi-\frac{1}{\xi}}, &\quad\mbox{if }$\xi\ne0$,\cr
\mu+\psi\log n,                &\quad\mbox{if }$\xi=0$.
}
\label{E3}
\end{equation}

Loosely, the $n$-year return value is the value that would be expected
to occur
once in $n$ years under a stationary climate. In this paper we take $n=100$,
though other values such as $n=25$ or $n=50$ could equally well be
taken. Return values allow one to summarize extreme precipitation in
one number, and are widely used and better understood by the common
practitioner than the GEV parameters. An alternative approach to
modeling the 100-year return value would be to model the three GEV
parameters separately as a function of spatial location. However, this
would introduce additional complications into the analysis (for
example, how to model the dependence among the three GEV parameters),
and we have preferred to use 100-year return values directly as this
leads to a simpler model.

\subsection{Threshold exceedances and the point process approach}\label{PP}

The simplest procedure for fitting the model (\ref{E1}) is to calculate
the annual maxima, say, $Y_1,\ldots,Y_M$, for a series of length $M$
years, and fit
(\ref{E1}) directly by maximum likelihood or some alternative statistical
technique. For example, the papers by Kharin and Zwiers (\citeyear{KharinZwiers2000}) and
Zwiers and Kharin (\citeyear{ZwiersKharin1998}) used the \textit{L-moments} technique which is popular among
hydrologists and meteorologists.

In the present context, however, the direct method has some
disadvantages. Fitting the GEV to annual maxima is problematic when
series are short. In addition, many of the series contain missing
values, and it is not clear how to adjust the annual maxima to
compensate for this.

Because of the difficulties associated with annual maxima, alternative methods
have become popular based on \textit{peaks over thresholds} (also known as the
POT approach). In this approach,
for each series a high threshold is selected, and a distribution fitted
to all
the values that exceed that threshold. Following Pickands (\citeyear{Pickands1975}), the
distribution
of exceedances over the threshold is taken to be the \textit{Generalized Pareto}
distribution (GPD), which asymptotically approximates the distribution of
exceedances over a threshold in the same sense as the GEV
asymptotically approximates
the distribution of maxima over a long time period. Davison and Smith (\citeyear{DavisonSmith1990})
developed a detailed statistical modeling strategy for exceedances over
thresholds
based on the GPD. Threshold-exceedance methods work better than annual-maxima
methods when the series is short, and also adapt themselves better to missing
values in the data.

For the present paper, however, we prefer a third approach, the
\textit{point process approach} [Smith (\citeyear{Smith1989,Smith2003}) and Coles (\citeyear{Coles2001})],
that, although operationally very similar to the POT approach, uses a
representation of the probability distribution that leads directly to
the GEV parameters $(\mu,\psi,\xi)$. An advantage of the point process
approach is that the parameter estimates are not directly tied to the
choice of threshold, and the ideal threshold can be determined by
considering where the parameter estimates stabilize. Although the
parameters for the point process approach are different from those of
the GPD approach, the two are still mathematically equivalent (in the
case used in the present paper, where there is no direct dependence on
covariates), so the consistency and asymptotic normality of estimators
follows from results in Smith (\citeyear{Smith1987}), among other references on
statistical properties for the GPD.

Under this model, if we observe $N$ peaks over a threshold $u$, say,
$Y_1,\ldots,Y_N$, at times $T_1,\ldots,T_N$, during an observational period
$[0,T]$, we view the pairs $(T_1,Y_1),\ldots,(T_N,Y_N)$ as points in the
space $[0,T] \times(u,\infty)$, which form a nonhomogeneous Poisson
process with intensity measure
%
\begin{equation}\label{E7}
\lambda(t,y) =
{1\over\psi} \biggl(1+\xi{y-\mu\over\psi} \biggr)_+^{-1/\xi-1}.
\end{equation}
The negative log-likelihood associated with this model may be written
in the form
%
\begin{eqnarray}\label{E9}
\ell(\mu,\psi,\xi)
&=& N\log\psi+ \biggl({1\over\xi}+1 \biggr)
\sum_{i=1}^N \log \biggl(1+\xi{Y_i-\mu\over\psi} \biggr)_+\nonumber
\\[-8pt]\\[-8pt]
&&{}+T \biggl(1+\xi{u-\mu\over\psi} \biggr)_+^{-1/\xi},\nonumber
\end{eqnarray}

\noindent
where $T$ is the length of the observation period in years and the
$(\cdots)_+$ symbols in (\ref{E9}) essentially mean that the expression is
evaluated only if $1+\xi{u-\mu\over\psi}>0$ and $1+\xi{Y_i-\mu\over\psi}>0$
for each $i$ (if these constraints are violated, we set $\ell=+\infty$).

The basic method of estimation is therefore to choose the parameters
$(\mu,\psi,\xi)$ to minimize (\ref{E9}). This is performed using
standard methods for numerical nonlinear optimization. Once we have
found the maximum likelihood estimates and associated
variance--covariance estimates, it is
straightforward to estimate the $n$-year return value from (\ref{E3}),
with an
approximation to the standard error of the estimate $\hat y_n$ by the
delta method.

\subsection{Details of the fitting procedure}\label{Details}

In practice, there are a number of details that need attention to
implement this procedure successfully:
\begin{enumerate}
\item
\textit{Missing values.} Missing values in the time series may be
accommodated by defining the time period $T$ in (\ref{E9}) to be the
total \textit{observed} time period, ignoring any periods when data are
missing. In practice, there could still be a bias if too many
observations are missing, because if there are trends in the data, the
results will be sensitive to the exact time period covered by the data.
To minimize this kind of bias, we impose the constraint that a station
is only included in the analysis if the proportion of missing days does
not exceed a small fraction $\varepsilon$, where, in practice, we take
$\varepsilon=0.1$.

\item
\textit{Seasonality.} Rainfall being a seasonal phenomenon, the GEV
parameters $(\mu,\psi,\xi)$ vary by season. Therefore, we perform
separate analyses for summer (June, July, and August), fall (September,
October, and November), winter (December, January, and February), and
spring (March, April, and May), where the calculation of $T$ in (\ref{E9}) is adjusted to account for the actual number of days in each
season (92 in summer, 91 in fall, 90.25 in winter allowing for leap
years, 92 in spring). We adopt the convention that December of each
year is counted as part of the following year, so there is not a
discontinuity in the winter season. For example, winter 1950 is
actually the period from December 1949 through Febuary 1950.

\item
\textit{Choice of time period.} After taking account of the convention
just noted regarding the month of December, the period over which
continuous records are available for both the point-source and gridded
data is 1949--1999. Our default option is therefore to take this as
defining the time period for our analysis. There may be some advantage
in considering shorter time periods, for example, to examine the extent
to which rainfall distributions have changed with time.

\item
\textit{Choice of threshold.} We follow the convention of taking a fixed
percentile at each station as the threshold for that station. For
example, the 95th-percentile threshold is defined as the 95th
percentile of all observations at a given station, excluding missing
values but including days when the observed precipitation is~0. As a
sensitivity check, we also considered the 97th percentile threshold but
find the results to be little different. It should be noted that papers
in the climate literature often consider much higher thresholds [e.g.,
Groisman et al. (\citeyear{GroismanEtAl2005}) use the 99.7\% threshold], but only in the
context of counting exceedances and not fitting probability
distributions to the excesses over a threshold. In the present context,
if we go much above the 97th percentile, we encounter too many failures
of the fitting algorithm.

\item
\textit{Clustering.} To compensate for short-term autocorrelations, we
usually work with \textit{peaks} over the threshold rather than all
individual exceedances over the threshold value, where the peaks are
defined as the largest values within each cluster. The \textit{runs
algorithm} [Smith and Weissman (\citeyear{SmithWeissman1994})] may be used to define peaks. In
practice, each group of consecutive daily observations over the
threshold was treated as a single cluster and only the cluster maximum
was used for the analysis. The results are not too sensitive to this
aspect of the analysis and, in fact, we would get very similar answers
if we treated every exceedance as a peak value.
\end{enumerate}

\subsection{Regression and model selection}

Once the return values were computed for both the point-level and
gridded data sets, each of the return values for the point-level (rain
gauge) data was identified with a particular grid box and the
associated return value for that grid box. A regression model was
fitted, using the gridded return values as a predictor for the
point-level return values.

The regression analysis considered the possibility of using transformations
or including additional covariates. It was found necessary to include
large-scale
spatial trends through polynomial functions of latitude and longitude.
Elevation was
also included in some of the regressions. A variety of strategies for
model selection
was adopted, including forward and backward selection, automatic model
fits through
Akaike's information criterion (AIC) [Akaike (\citeyear{Akaike1974})], and residual analysis.
Some of the analyses looked for spatial correlation among residuals
using the
variogram [Cressie (\citeyear{Cressie1993})] as a further diagnostic technique---an adequate
regression model should have spatially independent residuals. Details
of how these
analyses were conducted are in the next two sections.

\section{Results: Observational data (NCDC) versus reanalysis data (NCEP)}

In this section we detail results from the comparison between the
point-level (NCDC) 100-year return values versus the gridded reanalysis
(NCEP) 100-year return values. Figures labeled A.x are included in the
supplemental appendix.

\subsection{Extreme value analysis}

\begin{figure}[b]

\includegraphics{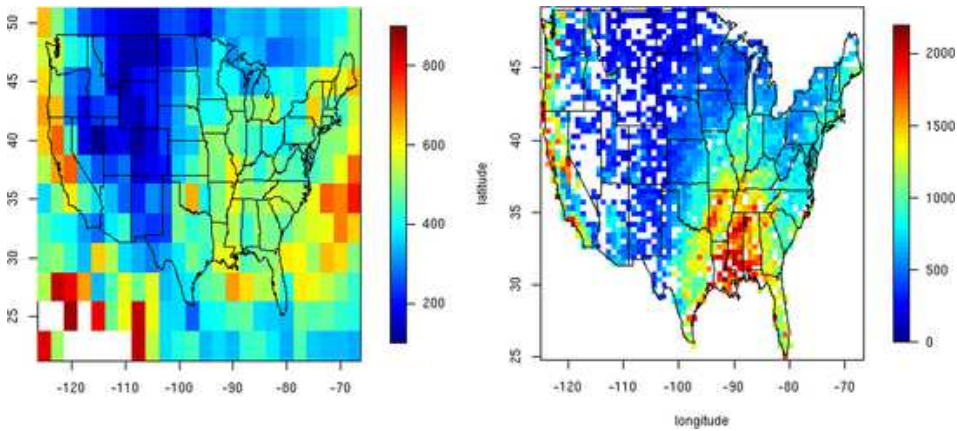}

\caption{\textup{NCEP grid and NCDC station return levels:}
One-hundred-year return values, in tenths of a millimeter, computed for
each grid cell in the NCEP data (left frame) and for each station from
the NCDC data (right frame).}\label{return}
\end{figure}

For an initial analysis, the focus is on the winter (DJF) season. The
GEV parameters were estimated via the point process approach as
described in Section~\ref{PP} for each station in the NCDC data and
each grid cell from the NCEP reanalysis data, using the 95th percentile
for each
data set as the threshold. As explained in Section~\ref{Details}, the
fitting method
allows for missing values, but we excluded stations with more than 10\% missing
values over the 1949--1999 time period. Of the original 5873 stations,
about 1530 were excluded by this criterion. In addition, for under 1\%
of all MLE calculations, the algorithm
failed to converge, and these were also excluded from subsequent analysis.
Figure~\ref{return} shows the 100-year return values computed for both
the NCDC station data and each grid cell from the NCEP data for the
winter season. The point location data (bottom frame) has a finer
plot-grid with one station-level return per grid. The sparseness (blank
grids) in the NCDC plot is due to some stations being excluded or
because some of the grid boxes had no stations originally.
%

The plots show the meteorological differences across the U.S. Two
aspects of these plots are immediately visible: (1) the spatial
patterns in the return values are very similar, with higher values in
the southeast and far west, and lower values over the upper mid-west
and mountain regions in the west, and (2) the return values for the
point-level data are much larger than those of the gridded data, as
evidenced by the difference in the scales of the two figures. The first
point is addressed using information available at each point station,
that is, latitude, longitude, and elevation measures.

\begin{figure}[b]

\includegraphics{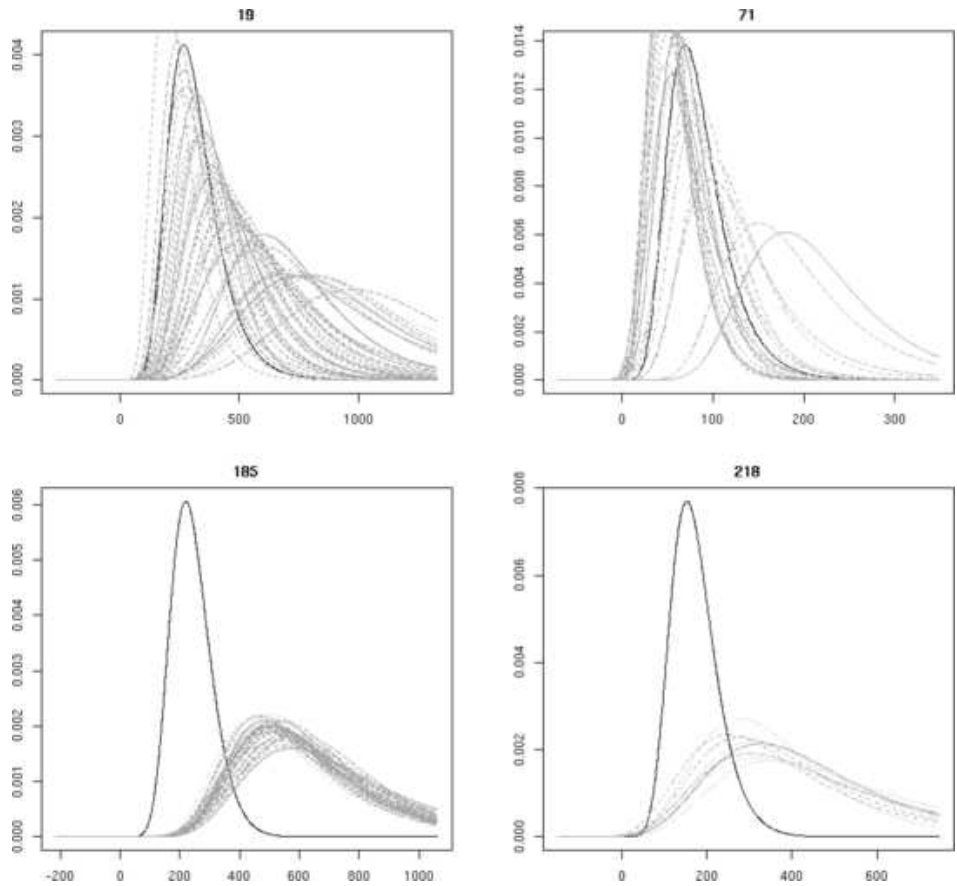}

\caption{\textup{Densities:}
Solid black curves indicate the fitted GEV densities for four grid
cells from the reanalysis (NCEP) grid data. San Francisco coast (top left):
Latitude 37.5, Longitude $-$122.5;
Montana (top right): Latitude 47.5, Longitude $-$112.5;
Alabama (bottom left): Latitdue 32.5, Longitude $-$87.5; and
Key Largo, FL (bottom right): Latitude 25, Longitude $-$80.
Dotted grey curves indicate the fitted densities from each NCDC station
within the grid cells.}\label{densities}
\end{figure}

Spatial trends are addressed through incorporation of local point
station measures. The last feature is detailed further in Figure~\ref{densities}, where the GEV parameter fits from the NCEP gridded data
(solid black curves) are compared to the GEV parameter fits from the
individual stations (dotted grey curves) within each grid cell. There
appears to be substantial variation among the density curves for
individual rain gauges, but the grid cell density seems clearly to be
from a different population for the grids from Alabama and Florida.
Thus, the impact of the aggregation across the grid cells is
immediately apparent. The point-level densities suggest larger return
values than those derived from the NCEP data in the corresponding grid
cells, which again is clearly seen for the grids in Alabama and Florida.

As explained in Section~\ref{Details}, a comparison was made between
the 95th and
97th percentile thresholds to examine the sensitivity of the analysis
to threshold
selection. GEV parameter estimates for these two threshold choices are
shown in Figures~A.1--A.3 [Mannshardt-Shamseldin et al. (\citeyear{Mannshardt-ShamEtAl2009})].
Comparing across the two thresholds, the GEV model parameters show
little change, suggesting that the parameter estimates have stabilized
and the 95th percentile is a high enough threshold. For subsequent
analysis, we use the 95th percentile for the threshold. It is generally
agreed upon in the literature that the shape parameter $\xi$ should be
small but positive.
For the station data, the mean $\hat{\xi}$ across all stations ranged
from $0.087$ in the spring to $0.127$ in the summer and the percentage
of stations for which $\hat{\xi}$ was positive ranged from $64\%$ in
the spring to $77\%$ in the fall. A $0.05$-level one-sided hypothesis
test for $\xi= 0$ was rejected 15--25\% of the time for the
right-sided alternative and 3--7\% for the left-sided alternative.
The results were similar for the grid cell (NCEP) data, however
somewhat less decisive. The average values of $\hat{\xi}$
over all the grid cells ranged from $0.04$ in the summer to $0.12$ in
the fall, however, in the spring, only $41\%$ of the $\hat{\xi}$
were positive, $17\%$ rejected $\xi= 0$ for a right-sided alternative
and $23\%$ for a left-sided alternative. These results do not
contradict that there is an overall tendency for $\xi$ to be positive,
but clearly there is a lot of variability from one data set to another.
This is not in conflict with the general belief that $\xi$ is greater
than zero, and we recognize that, in practice, there is so much
variability in individual estimates that it is difficult to make such
conclusions. The evidence is less decisive in the case of the grid cell
data than it is with the station data, which is consistent with our
concern that maybe NCEP does not represent extreme precipitation well.

\subsection{Direct comparison of NCEP and station averaged data}

To gain further insight into the relationship between extreme values in
NCEP and in station data, the following comparison was performed.

We selected 17 NCEP grid cells that included a large number ($>$65)
of observational stations. For each such grid cell, a ``station averaged''
data set was constructed by averaging precipitation values over all
stations on each day (including zeros, but omitting missing values).
The extreme value parameters were estimated for this station-averaged
data set and used to estimate the 100-year return value. The result (a) was
compared with (b) the 100-year return value estimated from NCEP data
and (c) the average of 100-year return values for each of the individual
stations in that grid cell. If NCEP data are an accurate representation,
we should expect (a) and (b) to be roughly comparable, but (c) to be
larger. Table~\ref{m1} shows the results for the DJF data.
%
\begin{table}
\caption{Table of 100-year return values computed for 17 grid
cells for the DJF season, by
\textup{(a)} first averaging daily station data,
the fitting an extreme value distribution to the daily averages;
\textup{(b)} fitting an extreme value distribution to the NCEP values;
\textup{(c)} averaging over
100-year return values computed for individual
stations}\label{m1}
\begin{tabular*}{\textwidth}{@{\extracolsep{\fill}}lcd{3.1}d{4.0}cd{4.0}@{}}
\hline
&
&
& \multicolumn{3}{c@{}}{\textbf{100-year return value estimated from}}
\\[-6pt]
& & & \multicolumn{3}{c@{}}{\hrulefill} \\
& \textbf{Latitude}
& \multicolumn{1}{c}{\textbf{Longitude}}
& \multicolumn{1}{c}{\textbf{Station averages}}
& \textbf{NCEP}
& \multicolumn{1}{c@{}}{\textbf{Averages of individual}} \\
  \multicolumn{1}{@{}c}{\textbf{Grid cell}}
& \multicolumn{1}{c}{$\bolds{(}^{\circ}\mathbf{N}\bolds{)}$}
& \multicolumn{1}{c}{$\bolds{(}^{\circ}\mathbf{W}\bolds{)}$}
& \multicolumn{1}{c}{$\bolds{(}\mathbf{a}\bolds{)}$}
& $\bolds{(}\mathbf{b}\bolds{)}$
& \multicolumn{1}{c@{}}{\textbf{station return values} $\bolds{(}\mathbf{c}\bolds{)}$} \\
\hline
\phantom{0}1 & 32.5 & 97.5 & 741 & 429 & 1170 \\
\phantom{0}2 & 32.5 & 95.0 & 1035 & 450 & 1531 \\
\phantom{0}3 & 32.5 & 90.0 & 1112 & 529 & 1691 \\
\phantom{0}4 & 32.5 & 85.0 & 954 & 505 & 1315 \\
\phantom{0}5 & 35.0 & 97.5 & 632 & 661 & 872 \\
\phantom{0}6 & 35.0 & 82.5 & 658 & 561 & 995 \\
\phantom{0}7 & 37.5 & 122.5 & 955 & 670 & 1559 \\
\phantom{0}8 & 37.5 & 100.0 & 300 & 351 & 488 \\
\phantom{0}9 & 37.5 & 97.5 & 498 & 438 & 669 \\
10 & 37.5 & 82.5 & 443 & 490 & 670 \\
11 & 37.5 & 80.0 & 446 & 452 & 713 \\
12 & 37.5 & 77.5 & 505 & 442 & 766 \\
13 & 40.0 & 97.5 & 307 & 360 & 545 \\
14 & 40.0 & 80.0 & 378 & 392 & 576 \\
15 & 40.0 & 77.5 & 526 & 451 & 758 \\
16 & 40.0 & 75.0 & 610 & 468 & 847 \\
17 & 42.5 & 90.0 & 300 & 493 & 489 \\
\hline
\end{tabular*}
\end{table}

In 10 of the 17 cases, the ratio of (a) to (b) is between 0.8 and 1.2.
Of the exceptions, the ratio is below 0.8 in one case (grid cell 17)
and ranges up to 2.3 (cell 2). In contrast, the ratio of (c) to (b) is
$>$1.3 in all but one case (cell 17), and goes as high as 3.4 in cell 2.
Similar results were obtained for the other three seasons; the ratios
overall [both (a) to (b) and (c) to (b)] were highest for the JJA season.

The results of this exercise show (as we anticipated) that NCEP data are
not an ideal representation of precipitation extremes computed from
averages over stations; but in a majority
of grid cells, the representation is reasonable, and it shows that the
discrepancy between return values computed from NCEP and from
individual stations is not primarily due to NCEP being a poor representation
of precipitation extremes.

\subsection{Regression results}

\begin{figure}

\includegraphics{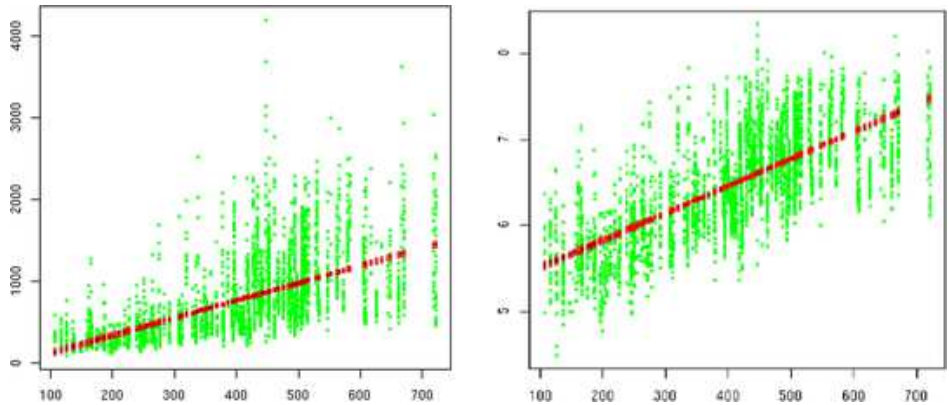}

\caption{\textup{Modeling point station return values:}
100-year return values: point-station returns regressed on gridded
return values (left) and log-transform of point-station values
regressed on gridded returns (right) for the NCEP grid cell data.
Return values are in tenths of a millimeter for the winter (DJF)
season.}\label{logtransform}
\end{figure}

Now that the return values have been computed, the discrepancy between
the return values computed from the NCDC point-level station data and
the NCEP gridded data can be modeled using regression methods. Each
point-level station is assigned to a grid cell and the relationship
between the grid cell return values and the point-level return values
is considered. A simple model, using return values from the grid cells
to predict the return values from the station data, shows excessive
dispersion. The dispersion is increasing for larger return values and
the model does little to capture the spatial trends nor account for the
differences in scale of the return values. However, an alternative
regression model using a logarithmic transformation of the point-level
return values greatly improves the fit of the regression model,
especially for homoscedasticity. This can be seen in Figure~\ref{logtransform}. A model where both the grid cell returns and the
station level returns were log-transformed was considered, however,
there was no significant difference in the fit of the model with both
transformed. The AIC value of the model with just the station level
returns log-transformed was smaller than the model with both
transformed, and a comparison of standard errors showed that the
log--log model had relatively higher standard errors than the model with
only the station level returns transformed. Thus, the model where just
the station level returns were log-transformed was chosen for further analysis.
Adding elevation, measured in meters, as a covariate also reduced the
dispersion. Details of the model fits for all four seasons are shown in
Table~\ref{m2}.
%
\begin{table}
\caption{Model coefficients across season and threshold
for the basic regression model, log(Point return levels) $\sim$ Grid
Return $+$ Elevation (no latitude or longitude terms), for the NCEP grid
data. All coefficient $p$-values $< 0.001$. Standard error are based on
the assumption of independence, which is further shown to be possibly
an invalid assumption}\label{m2}
\begin{tabular*}{\textwidth}{@{\extracolsep{\fill}}lcccccc@{}}
\hline
 & \textbf{Int.} & \textbf{SE} & \textbf{Grid} & \textbf{SE} & \textbf{Elev} & \textbf{SE} \\
\hline
Winter\\
95th & 5.32 & 0.0032& 0.0030 & 6.0$\times10^{-5}$& $-$0.00012 &1.5$\times10^{-5}$ \\
97th & 5.31 & 0.0030 & 0.0030 & 5.8$\times10^{-5}$ & $-$0.00011 &1.5$\times10^{-5}$ \\
Spring & & & & & & \\
95th & 5.90 & 0.029 & 0.0023 & 5.6$\times10^{-5}$ & $-$0.00026 &1.4$\times10^{-5}$ \\
97th & 5.97 & 0.029 & 0.0021 & 5.7$\times10^{-5}$ & $-$0.00028 &1.1$\times10^{-5}$ \\
Summer & & & & & & \\
95th & 6.65 & 0.027 & 0.0015 & 6.1$\times10^{-5}$ & $-$0.00039 &9.3$\times10^{-6}$ \\
97th & 6.58 & 0.027 & 0.0016 & 6.0$\times10^{-5}$ & $-$0.00037 &9.6$\times10^{-6}$ \\
Fall & & & & & & \\
95th &6.93 & 0.026 & 0.0006 & 4.4$\times10^{-5}$ & $-$0.00049 &1.1$\times10^{-5}$ \\
97th & 6.83 & 0.026 & 0.0008 & 4.4$\times10^{-5}$ & $-$0.00048 &1.2$\times10^{-5}$ \\
\hline
\end{tabular*}
\end{table}

There does not appear to be much difference in the fitted models for
the different threshold values, supporting the previous conclusion
concerning the GEV model parameters. There does, however, appear to be
a seasonal effect, with the coefficients on both the grid return values
and elevation changing between the seasons. Both of the covariates are
significant in all of the models. In subsequent analysis we
concentrate on the winter season (DJF) in order to investigate scaling
and spatial trends.

\subsection{Spatial trends}

\begin{figure}[b]

\includegraphics{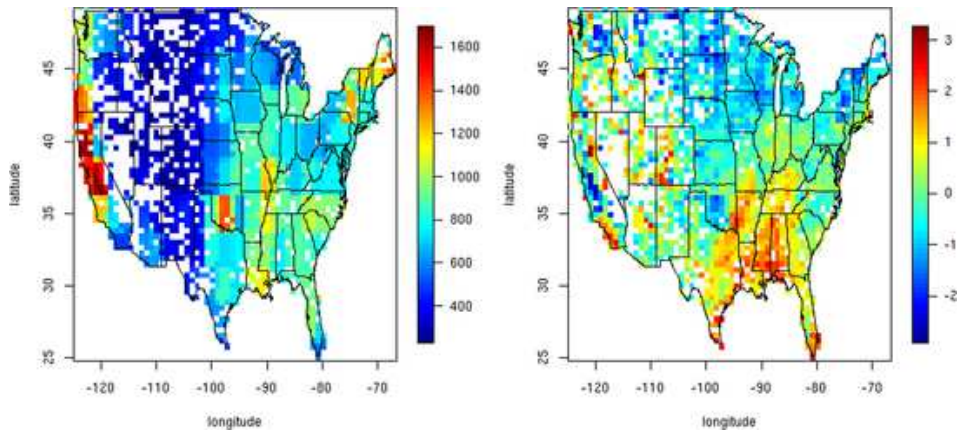}

\caption{\textup{Simple regression model including elevation:}
Fitted return values (left) and residuals (right) from the regression
model Log(Point) $\sim$ Grid $+$ Elevation using the NCEP grid cell
data.}\label{resids1}
\end{figure}

The simple regression of log point-level return value on grid return
value appears to do an adequate job of predicting the return values for
the station data, but the residuals still show spatial patterns that
are not being accounted for (see Figure~\ref{resids1}). In particular,
the right plot in Figure~\ref{resids1} shows generally positive
residuals in the southeast, and generally negative residuals in several
other regions (midwest, northeast, west coast). Such clear spatial
patterns are indicative that the residuals are nonrandom and,
therefore, further modeling is required.

\begin{figure}

\includegraphics{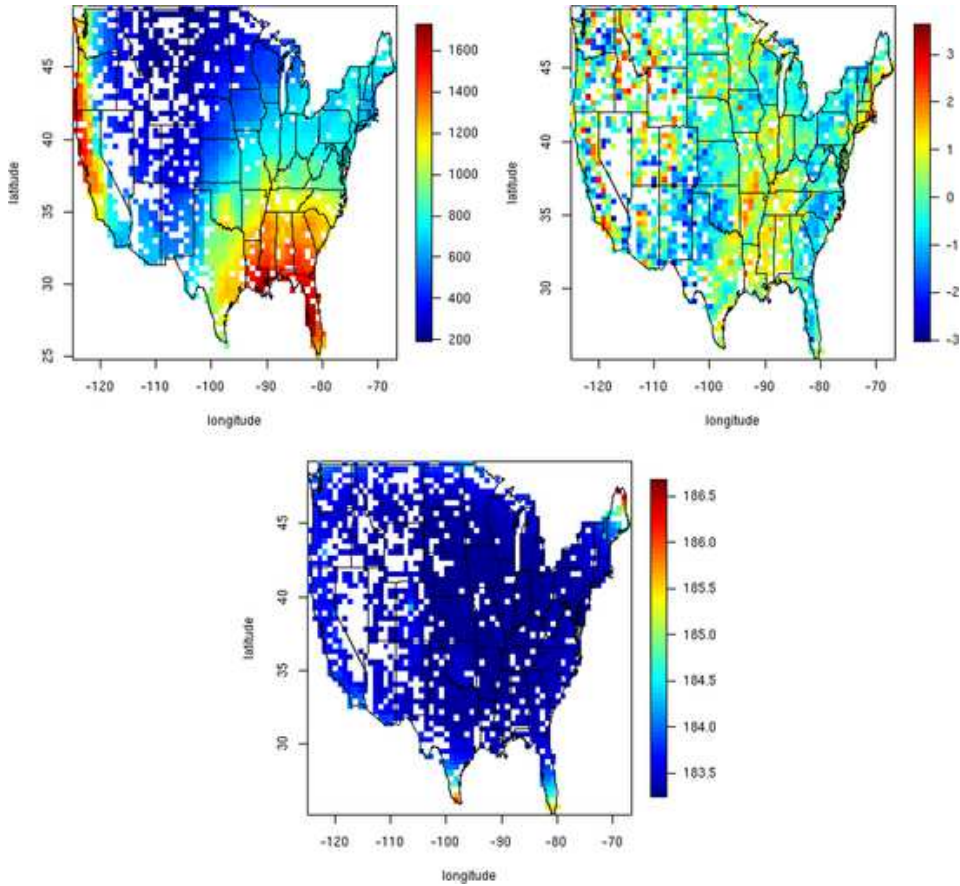}

\caption{\textup{Cubic model in lat and lon including elevation:}
Fitted return values (top left), residuals (top right), and prediction
standard errors (bottom) from the cubic regression model with NCEP grid
data. Errors range between 183.2 and 186.7 tenths of a mm.}\label{resids2}
\vspace*{-4pt}
\end{figure}
Therefore, the model was extended to include polynomial terms in
latitude and longitude. The cubic model appears to do the best job at
capturing the spatial trends (Figure~\ref{resids2}). Although the
residual plot from the cubic model (right plot, Figure~\ref{resids2})
still shows some evidence of spatial dependence, it is not nearly as
strong as the corresponding residual plot in Figure~\ref{resids1} and,
in fact, subsequent tests imply it may be spurious. Empirical
variograms [shown in Figure~A.4, Mannshardt-Shamseldin et al. (\citeyear{Mannshardt-ShamEtAl2009})]
indicate a lack of spatial dependence in the residuals from the cubic
model. The variogram is a measure of the variability between two
observations at a given distance apart. The essentially flat variogram
for the cubic model, shown in Figure~A.4, indicates little spatial
dependence, whereas the gradual rise toward the maximum, as seen in the
plot for the linear model with no latitude or longitude terms, is
indicative of spatial dependence. The residuals for the cubic model,
Figure~\ref{resids1}, show some evidence of spatial dependence for the
eastern half of the U.S. Motivated by the appearance of a trend among
sites in the eastern United States seen in the residuals of Figure~\ref{resids2}, at the suggestion of a reviewer, the bottom variogram in
Figure~A.4 details the spatial trend for stations east of 100$^{\circ}$W.
The essentially flat variogram for the cubic model indicates little
spatial dependence.
Additionally, all of the main effects and interactions in longitude and
latitude were significant in all four seasons.

We also considered higher-order models, in particular, one including
quartic terms in latitude and longitude. For all four seasons, the
quartic model is superior to the cubic model based on AIC. However,
from other points of view the cubic model seems superior. There is
little difference in the predictions based on the two models, and
several terms in the quartic model (as well as elevation) were not
significant. Individual terms in the quartic model are much harder to
interpret. The residuals from the cubic regression appear uncorrelated
based on the variogram plot (Figure~A.4), and a later exercise that
made a direct comparison with kriging for a subset of rain gauge sites
(Section~\ref{KrigCompareSection}) suggested that the cubic regression
method performed as well as kriging. Therefore, we did not pursue any
regression model beyond the one that included cubic terms in latitude
and longitude. There is no perfect model---the residuals from the
cubic model still show some
evidence of a trend, but it is enormously improved over the model with
no latitude and longitude terms.

\section{Results: Observational data (NCDC) versus climate model data\break (CCSM)}

The results of the previous section show that the extreme behavior on
gridded data sets can be used to model and predict extreme behavior at
specific point-level locations. However, the ultimate objective is to
apply the results to future projections from a climate model to obtain
projection of return values for point-level precipitation. Therefore,
we apply the same methodology to output from the CCSM model runs for
the winter season of the current time period 1970--1999 and a future
model run for 2070--2099. It is recognized that in contrast to the NCEP
analysis, the CCSM model is not constrained by observed weather
variables and therefore is not expected to reproduce the observed
precipitation as well as the NCEP reanalysis. In general, the parameter
estimates behave similarly to those based on the NCEP data. The spatial
patterns of parameters and return values of the NCDC point-level data
are consistent with the spatial patterns of the CCSM for the current
time period. Again, a cubic model is used to relate the return values
based on the gridded CCSM output to the NCDC point-level data. The
relationship of the CCSM grid-level return values to NCDC point-level
return values is similar to the relationship derived from NCEP/NCDC
data. The standard errors of the predicted values for the CCSM
1970--1999 model runs range between 203.9 and 205.5 tenths of a
millimeter for predicted values between 179.9 and 2016 tenths of a
millimeter. The standard errors of the predicted values are between
212.8 and 216.2 tenths of a millimeter when the same regression
relationship is applied to the CCSM 2070--2099 model runs, where
the
predicted values range from 173.7 and 2318 tenths of a millimeter. This
is comparable to the standard errors for the NCEP predicted values,
which ranged between 183.2 and 186.7 tenths of a millimeter for
predicted values between 194.0 and 1697 tenths of a millimeter.

\subsection{Kriging comparison}\label{KrigCompareSection}

Approximately 1530 stations out of 5873 were not used in the regression
analysis due to not meeting the 10\% missing criterion or
nonconvergence of the numerical algorithm used in estimating the GEV
parameters. To investigate the effectiveness of accounting for the
spatial dependence between stations, universal kriging is performed to
obtain predictions at the unused sites using the site latitude,
longitude, and elevation values. The kriging is performed using the R
function Krig from the fields package [Fields Development Team (\citeyear{FieldsDevTeam2006})], using an
exponential covariance structure with range parameter 155 miles [based
on mid-level range of 250~km used by Groisman et al. (\citeyear{GroismanEtAl2005})]. The kriged
values are compared to the predictions obtained through applying the
cubic model.

The extreme precipitation levels at the unused stations are very
similar for the cubic model predictions and kriged predictions [see
Figure~A.5, Mannshardt-Shamseldin et al. (\citeyear{Mannshardt-ShamEtAl2009})]. This suggests that
the cubic model has accounted for the spatial correlation between
stations with similar effectiveness as a more costly run-time kriging
analysis. Looking across just the unused stations at the ratio of
kriged to cubic model predictions, it can be seen that the kriged and
modeled return levels produce very similar predicted values.

\subsection{Future vs present CCSM returns}

\begin{figure}[b]

\includegraphics{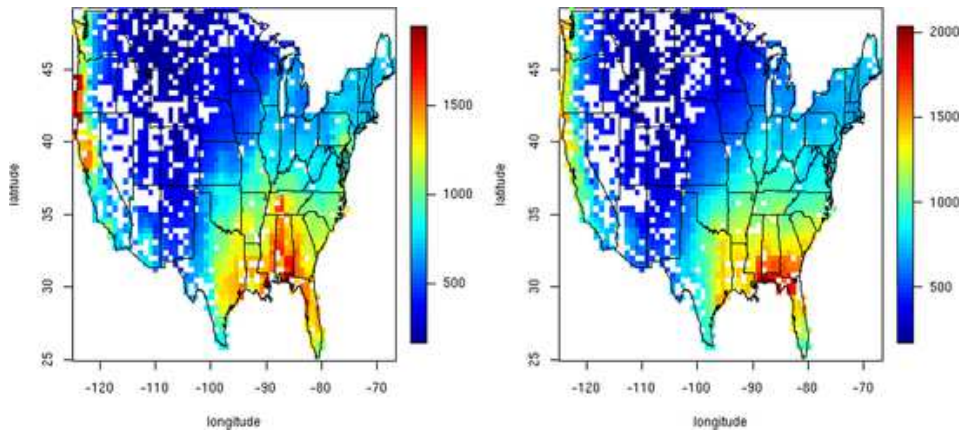}

\caption{\textup{Present and future return levels:}
Cubic Models of CCSM: Present return values for the current time period 1970--1999,
Max${}={}$2016 (left plot); and Future return values for the time period 2070--2099,
Max${}={}$2318 (right plot).}\label{future}
\vspace*{-2pt}
\end{figure}
The spatial pattern is consistent across present and future model run
predictions of 100-year return values. However, the scale is different---an
increased scale is seen for the future predictions in Figure~\ref{future}. The standard errors for the ratio were computed using the
delta method and range between $0.017$ and $3.241$ tenths of a
millimeter. It should be noted that a few of the standard errors for
the ratio seemed unrealistically large, but 95\% were less than 0.444
tenths of a millimeter.

\begin{figure}

\includegraphics{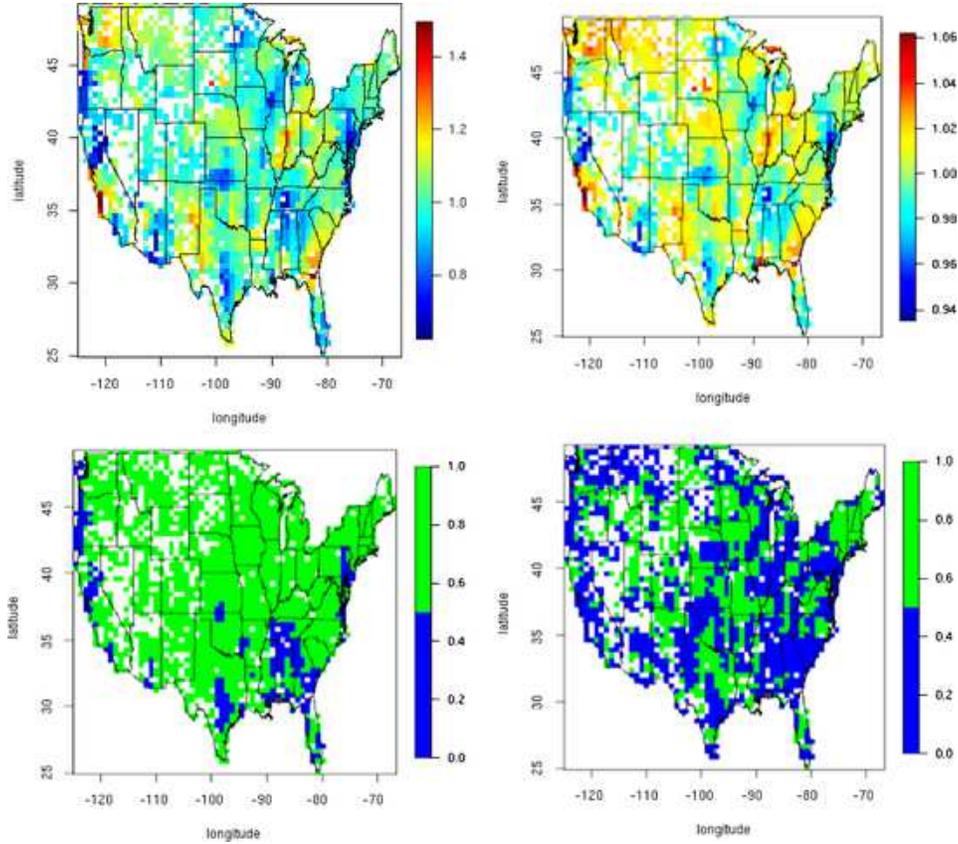}

\caption{\textup{Ratio future to present return levels:}
Ratio of predicted point-level return values for the future run
(2070--2099) to the predicted point-level return values for the control
run (1970--1999) for the CCSM grid cell data (top left). Ratio of the
cubic model outputs (top right).
Indicator of stations where ratio is significantly different from 1.0
(0.05 level). Bottom left shows the indicators of the scaled return
value ratios, bottom right shows indicators of log-scale ratio. ``1''
(green) indicates no significant difference, ``0'' (blue) indicates
ratio is significantly different from 1.0.}\label{ccsm}
\end{figure}

We calculate the ratio of predicted point-location return values for
the present-day run to the predicted return values for the future run.
Figure~\ref{ccsm} displays these results. Figure~\ref{ccsm} also shows
the ratio of predicted point-level return values for the future run to
the predicted point-level return values for the control run for
log-scale model outputs (top right).
The standard errors of the ratios on both the return-level scale and
the log-scale show a strong spatial trend and indicate the highest
levels of variability in the Mid-Northwest.
Generally, there is consistency between the return values, that
is,~many of the ratios are near 1.0. However, indicated are several
coastal areas where future predictions are up to twice the magnitude of
the current predictions, including the Southeast Coast and the Pacific
Northwest. In contrast, in-land areas of the South East, East North
Central, and Central regions show a predicted decrease.
In order to assess the significance of the increase or decrease in
return values suggested by the ratios, prediction intervals are
calculated. The bottom plots of Figure~\ref{ccsm} display an indicator
of stations where ratio is significantly different from 1.0 at the 0.05
level. Note that a significant difference from 1.0 is seen in the
Northwest coast, an area of the east coast, an area in southern Texas,
and the Southeast regions---which indicated an increase or decrease in
return values in the top plots of Figure~\ref{ccsm}.

\section{Discussion}

The analysis in this paper shows that for rain gauge and climate model
precipitation extremes, modeling the tail of the GEV distribution
produces stable GEV parameter estimates and model coefficients within
seasons and across 95\% and 97\% thresholds. In addition, we were able
to find regression relationships between the rain gauge (station-level)
and climate model (grid-level) extremes. 100-year return values are
successfully modeled by season at the point (station) level using
grid-level return values, station elevation, and station latitude and
longitude coordinates. For both the NCEP and CCSM return values, the
regression relationship between return values based on gridded and
point-location data is best expressed through a cubic model in latitude
and longitude. This in turn allows us to compute projected future
return values for point-location data based on the output of a climate model.

There is evidence of increasing extremes over time, as seen in the CCSM
grid cell data along several coastal areas where the future predictions
are up to two times the magnitude of the current modeled precipitation extremes.

The advantage of the present approach is its simplicity, requiring no
more than a combination of two well-established statistical techniques,
GEV analysis to calculate the return values and regression to relate
the point-location and gridded values. An alternative approach would be
to proceed more directly through spatial models of daily precipitation
fields. Coles and Tawn (\citeyear{ColesTawn1996}) developed such a relationship using
max-stable processes, which are especially appropriate in the context
of extremes. Later Sans\'o and Guenni (\citeyear{SansoGuenni2000,SansoGuenni2004}) derived a spatial
model for precipitation using a thresholded Gaussian process to
accommodate the fact that precipitation data includes both zero and
nonzero values. Although the Sans\'o--Guenni papers were not focussed
specifically on extremes, it is possible that their models, or some
variant of them, could also effectively explain the spatial patterns of
extremes. Our major reason for not pursuing these approaches here is
that they would require much more intense computations to be applied to
such a large data set as the entire precipitation record of the
continental United States. Nevertheless, we believe that attempting to
unify our present approach with one based on a stochastic model for
precipitation is an important topic for future work.

\section{Acknowledgments}

This research is supported in part through a grant from NCAR's Weather
and Climate Impact Assessment Science Program,
through the National
Science Foundation support of the Geophysical Statistics Program at
NCAR (DMS-03-55474), and in part by NOAA grant ``Statistical Assessment
of Uncertainty in Present and Future North American Rainfall Extremes''
(Richard L. Smith and Gabriele Hegerl, co-PIs). This material was
based upon work supported by the National Science Foundation under
Agreement No. DMS-06-35449. Any opinions, findings, and conclusions or
recommendations expressed in this material are those of the author(s)
and do not necessarily reflect the views of the National Science Foundation.

\begin{supplement}[id=suppA]
\stitle{Appendix of Graphics}
\slink[doi]{10.1214/09-AOAS287SUPP}
\slink[url]{http://lib.stat.cmu.edu/aoas/287/supplement.pdf}
\sdatatype{.pdf}
\sdescription{The appendix provides supplemental graphics, which are
referenced in the manuscript as~A.x.}
\end{supplement}

\printaddresses


\begin{thebibliography}{99}

\bibitem[\protect\citeauthoryear{}{1974}]{Akaike1974}
\textsc{Akaike, H.} (1974).
A new look at statistical model identification.
\textit{IEEE Trans. Autom. Control} \textbf{AU-19} 716--722.
\MR{0423716}

\bibitem[\protect\citeauthoryear{}{2001}]{Coles2001}
\textsc{Coles, S.~G.} (2001).
\textit{An Introduction to Statistical Modeling of Extreme Values}.
Springer, New York.
\MR{1932132}

\bibitem[\protect\citeauthoryear{}{1996}]{ColesTawn1996}
\textsc{Coles, S.~G.} and \textsc{Tawn, J.~A.} (1996).
Modelling extremes of the areal rainfall process.
\textit{J. Roy. Statist. Soc. Ser.~B} \textbf{58} 329--347.
\MR{1377836}

\bibitem[\protect\citeauthoryear{}{1993}]{Cressie1993}
\textsc{Cressie, N.} (1993).
\textit{Statistics for Spatial Data}, rev.~ed.
Wiley, New York.
\MR{1239641}

\bibitem[\protect\citeauthoryear{}{1990}]{DavisonSmith1990}
\textsc{Davison, A.~C.} and \textsc{Smith, R.~L.} (1990).
Models for exceedances over high thresholds (with discussion).
\textit{J. Roy. Statist. Soc. Ser. B} \textbf{52} 393--442.
\MR{1086795}

\bibitem[\protect\citeauthoryear{}{2006}]{FieldsDevTeam2006}
\textsc{Fields Development Team} (2006).
Fields: Tools for spatial data.
National Center for Atmospheric Research, Boulder, CO.
Available at
\href{http://www.image.ucar.edu/GSP/Software/Fields}{http://www.image.ucar.edu/GSP/}
\href{http://www.image.ucar.edu/GSP/Software/Fields}{Software/Fields}.

\bibitem[\protect\citeauthoryear{}{1928}]{FisherTippett1928}
\textsc{Fisher, R.~A.} and \textsc{Tippett, L.~H.~C.} (1928).
Limiting forms of the frequency distributions of the largest or
smallest member of a sample.
\textit{Proc. Camb. Phil. Soc.} \textbf{24} 180--190.

\bibitem[\protect\citeauthoryear{}{2002}]{FrichEtAl2002}
\textsc{Frich, P., Alexander, L.~V., Della-Marta, P., Gleason, B.,
Haylock, M., Klein Tank, A. M. G.} and \textsc{Peterson, T.} (2002).
Observed coherent changes in climatic extremes during the second half
of the twentieth century.
\textit{Climate Res.} \textbf{19} 193--212.

\bibitem[\protect\citeauthoryear{}{1999}]{GroismanEtAl1999}
\textsc{Groisman, P. Y., Karl, T.~R., Easterling, D.~R.,
Knight, R. W., Jamason, P. F., Hnnessy, K. J., Suppiah, C.,
Wibig, J., Fortuniak, K., Razuvaev, N. V., Douglas, A., F\o rland, E.} and \textsc{Xhai, P.-M.} (1999).
Changes in the probability of heavy precipitation: Important indicators
of climatic change.
\textit{Climatic Change} \textbf{42} 243--283.

\bibitem[\protect\citeauthoryear{}{2005}]{GroismanEtAl2005}
\textsc{Groisman, P.~Y., Knight, R.~W., Easterling, D.~R.,
Karl, T. R., Hegerl, G. C.} and \textsc{Razuvaev, V. N.} (2005).
Trends in intense precipitation in the climate record.
\textit{Journal of Climate} \textbf{18} 1326--1350.

\bibitem[\protect\citeauthoryear{}{1958}]{Gumbel1958}
\textsc{Gumbel, E.~J.} (1958).
\textit{Statistics of Extremes}. Columbia Univ. Press, Cambridge.
\MR{0096342}

\bibitem[\protect\citeauthoryear{}{2004}]{HegerlEtAl2004}
\textsc{Hegerl, G.~C., Zwiers, F.~W., Stott, P.~A.,
Kanamitsu, M., Kistler, R., Collins, W., Deaven, D., Gandin, L.,
Iredell, M., Saha, S., White, G., Woollen, J., Zhu, Y., Leetmaa, A.} and \textsc{Reynolds, R.} (2004).
Detectability of anthropogenic changes in annual temperature and
precipitation extremes.
\textit{Journal of Climate} \textbf{17} 3683--3700.

\bibitem[\protect\citeauthoryear{}{1996}]{KalnayEtAl1996}
\textsc{Kalnay E.} (1996).
The NCEP/NCAR 40-year reanalysis project.
\textit{Bull. Amer. Meteor. Soc.} \textbf{77} 437--470.

\bibitem[\protect\citeauthoryear{}{1998}]{KarlKnight1998}
\textsc{Karl, T.~R.} and \textsc{Knight, R. W. }(1998).
Secular trends of precipitation amount, frequency, and intensity in the
United States.
\textit{Bull. Am. Met. Soc.} \textbf{79} 231--241.

\bibitem[\protect\citeauthoryear{}{2002}]{KatzParlangeNaveau2002}
\textsc{Katz, R.~W., Parlange, M.~B.} and \textsc{Naveau, P.} (2002).
Statistics of extremes in hydrology.
\textit{Advances in Water Resources} \textbf{25} 1287--1304.

\bibitem[\protect\citeauthoryear{}{2000}]{KharinZwiers2000}
\textsc{Kharin, V. V.} and \textsc{Zwiers, F.~W.} (2000).
Changes in the extremes in an ensemble of transient climate simulations
with a coupled atmosphere-ocean GCM.
\textit{Journal of Climate} \textbf{13} 3760--3788.

\bibitem[\protect\citeauthoryear{}{2003}]{KiktevEtAl2003}
\textsc{Kiktev, D., Sexton, D., Alexander, L.} and \textsc{Folland, C.} (2003).
Comparison of modeled and observed trends in indices of daily climate extremes.
\textit{Journal of Climate} \textbf{16} 3560--3571.


\bibitem[\protect\citeauthoryear{}{2010}]{Mannshardt-ShamEtAl2009}
\textsc{Mannshardt-Shamseldin, E. C., Smith, R. L., Sain, S., Mearns, L. O.} and \textsc{Cooley, D.} (2010).
Supplement to ``Downscaling extremes: A comparison of extreme value
distributions in point-source and gridded precipitation data.''
DOI:
\href{http://dx.doi.org/10.1214/09-AOAS287SUPP}{10.1214/09-AOAS287SUPP}.

\bibitem[\protect\citeauthoryear{}{2000}]{MeehlEtAl2000}
\textsc{Meehl, G.~A., Zwiers, F., Evans, J., Knutson, T., Mearns, L.} and \textsc{Whetton, P.} (2000).
Trends in extreme weather and climate events: Issues related to
modeling extremes in projections of future climate change.
\textit{Bull. Am. Met. Soc.} \textbf{81} 427--436.

\bibitem[\protect\citeauthoryear{}{1975}]{Pickands1975}
\textsc{Pickands, J.} (1975).
Statistical inference using extreme order statistics.
\textit{Ann. Statist.} \textbf{3} 119--131.
\MR{0423667}

\bibitem[\protect\citeauthoryear{}{2000}]{SansoGuenni2000}
\textsc{Sans\'o, B.} and \textsc{Guenni, L.} (2000).
A non-stationary multisite model for rainfall.
\textit{J. Amer. Statist. Assoc.} \textbf{95} 1064--1089.
\MR{1821717}

\bibitem[\protect\citeauthoryear{}{2004}]{SansoGuenni2004}
\textsc{Sans\'o, B.} and \textsc{Guenni, L.} (2004).
A Bayesian approach to compare observed rainfall data to deterministic
simulations.
\textit{Environmetrics} \textbf{15} 597--612.

\bibitem[\protect\citeauthoryear{}{1987}]{Smith1987}
\textsc{Smith, R.~L.} (1987).
Estimating tails of probability distributions.
\textit{Ann. Statist.} \textbf{15} 1174--1207.
\MR{0902252}

\bibitem[\protect\citeauthoryear{}{1989}]{Smith1989}
\textsc{Smith, R.~L.} (1989).
Extreme value analysis of environmental time series: An application to
trend detection in ground-level ozone (with discussion).
\textit{Statist. Sci.} \textbf{4} 367--393.
\MR{1041763}


\bibitem[\protect\citeauthoryear{}{2003}]{Smith2003}
\textsc{Smith, R.~L.} (2003).
Statistics of extremes, with applications in environment, insurance and finance.
In \textit{Extreme Values in Finance, Telecommunications and the Environment}
(B. Finkenstadt and H.~Rootzen, eds.). Chapman and Hall/CRC Press, London.

\bibitem[\protect\citeauthoryear{}{1994}]{SmithWeissman1994}
\textsc{Smith, R.~L.} and \textsc{Weissman, I.} (1994).
Estimating the extremal index.
\textit{J. Roy. Statist. Soc. Ser.~B} \textbf{56} 515--528.
\MR{1278224}


\bibitem[\protect\citeauthoryear{}{1998}]{ZwiersKharin1998}
\textsc{Zwiers, F.~W.} and \textsc{Kharin, V.~V.} (1998).
Changes in the extremes of the climate simulated by CCC GCM2 under
CO$_2$ doubling.
\textit{Journal of Climate} \textbf{11} 2200--2222.
\end{thebibliography}
\end{document}